\documentclass{article}

\usepackage[preprint]{neurips_2026}
\usepackage{graphicx}
\usepackage{tikz}
\usepackage[absolute,overlay]{textpos}
\usepackage{tcolorbox}
\usepackage{array}
\usepackage{multirow}
\usepackage{xcolor}
\usepackage{fontawesome5}
\usepackage{hyperref} 
\usepackage{fontawesome5}

\definecolor{academicblue}{RGB}{0, 51, 102} 

\AddToHook{shipout/before}{%
  \pdfpageattr{/Group<</S/Transparency /I true /CS/DeviceRGB>>}%
}

\usepackage[table]{xcolor} 
\usepackage{booktabs}
\usepackage{graphicx}

\definecolor{tabletint}{RGB}{242, 246, 250} 
\definecolor{academicblue}{RGB}{0, 51, 153} 

\usepackage[utf8]{inputenc}
\usepackage[T1]{fontenc}
\usepackage{microtype}

\usepackage{hyperref}
\usepackage{url}

\usepackage{amsfonts}
\usepackage{nicefrac}
\usepackage{siunitx}

\usepackage{booktabs}
\usepackage{multirow}
\usepackage{makecell}
\usepackage{threeparttable}
\usepackage{adjustbox}
\usepackage{tabularx}
\usepackage{graphicx}
\usepackage{array}
\usepackage[table]{xcolor}
\usepackage{colortbl}
\usepackage{soul}
\usepackage{tikz}
\usepackage{amsmath}
\usepackage[table]{xcolor}
\usepackage{amsmath, amsthm, amsfonts}

\usepackage{empheq}
\usepackage{tcolorbox}
\usepackage{titletoc}
\usepackage{algorithmic}
\usepackage{algorithm}

\newcolumntype{Y}{>{\centering\arraybackslash}X}

\definecolor{lightgray}{gray}{0.95}
\definecolor{headbg}{RGB}{245,247,250}
\definecolor{subheadbg}{RGB}{250,251,253}
\definecolor{groupbg}{RGB}{248,249,251}
\definecolor{nacol}{RGB}{150,150,150}
\definecolor{bestgray}{gray}{0.90}
\definecolor{best}{RGB}{198,239,206}
\definecolor{second}{RGB}{226,239,218}
\definecolor{head}{RGB}{245,246,250}
\definecolor{subhead}{RGB}{250,250,252}
\definecolor{light}{RGB}{248,249,251}

\definecolor{pastelorange}{RGB}{232,150,75}
\definecolor{pastelblue}{RGB}{70,145,210}

\tcbuselibrary{skins}
\usepackage{titlesec}
\usepackage[labelfont=bf, font=small, labelsep=period]{caption}
\usepackage[absolute,overlay]{textpos}
\usepackage{microtype} 

\newcommand{\qbdisplay}[1]{\[\adjustbox{max width=\columnwidth}{$\displaystyle #1$}\]}

\newcommand{\qbsepn}{\quad}
\newcommand{\qbbreak}{\\}

\newcommand{\qblead}[1]{\textbf{#1}}
\newcommand{\qbemph}[1]{#1}

\DeclareUnicodeCharacter{2212}{\ensuremath{-} }            
\DeclareUnicodeCharacter{2191}{\ensuremath{\uparrow}}     
\DeclareUnicodeCharacter{2193}{\ensuremath{\downarrow}}   
\DeclareUnicodeCharacter{2192}{\ensuremath{\rightarrow}}  
\DeclareUnicodeCharacter{0394}{\ensuremath{\Delta}}       
\DeclareUnicodeCharacter{00B1}{\ensuremath{\pm}}          

\definecolor{DeepIndigo}{RGB}{10, 20, 100}
\definecolor{ElectricCyan}{RGB}{0, 220, 255}
\definecolor{NeonYellow}{RGB}{255, 230, 0}
\definecolor{VividRed}{RGB}{220, 20, 10}
\definecolor{DarkCrimson}{RGB}{110, 0, 5}
\definecolor{caspianbg}{RGB}{248, 250, 255} 

\hypersetup{
    colorlinks=true,
    linkcolor=DeepIndigo,
    citecolor=DeepIndigo,
    urlcolor=ElectricCyan
}

\titleformat{\section}
  {\normalfont\Large\bfseries\color{DeepIndigo}}{\thesection}{1em}{}
\titleformat{\subsection}
  {\normalfont\large\bfseries\color{DeepIndigo!80}}{\thesubsection}{1em}{}

\begin{document}

\begin{center}
    \vspace*{-1.2cm}
    \begin{minipage}[c]{0.15\linewidth}
        \includegraphics[height=0.85cm]{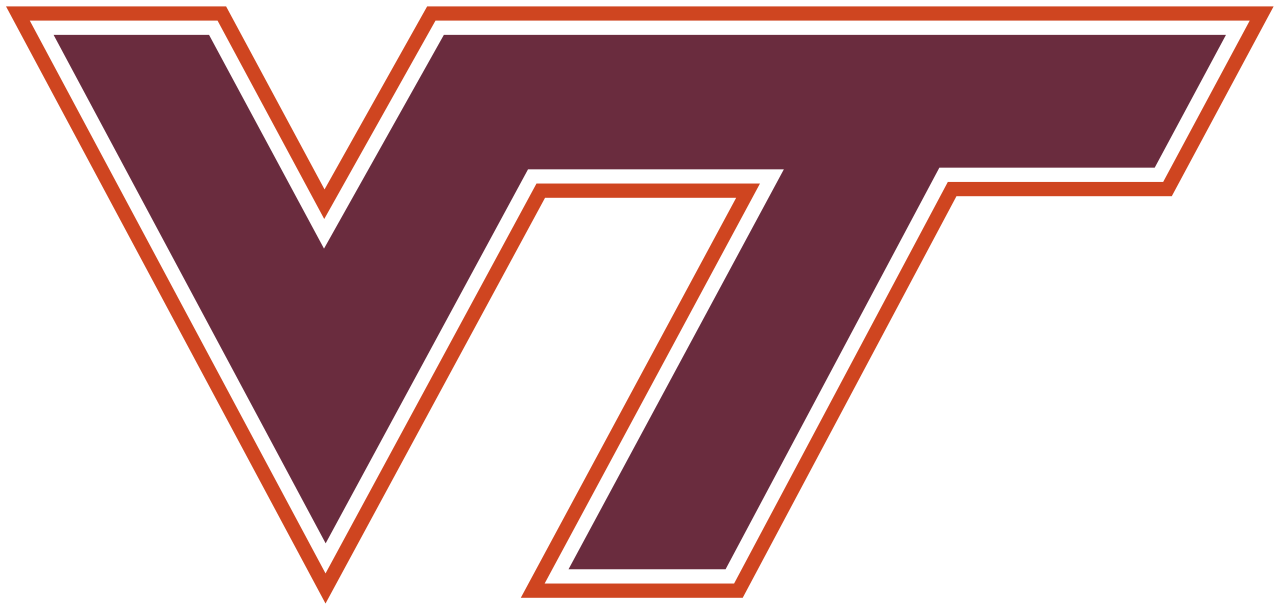}
    \end{minipage}%
    \begin{minipage}[c]{0.15\linewidth}
        \includegraphics[height=1.1cm]{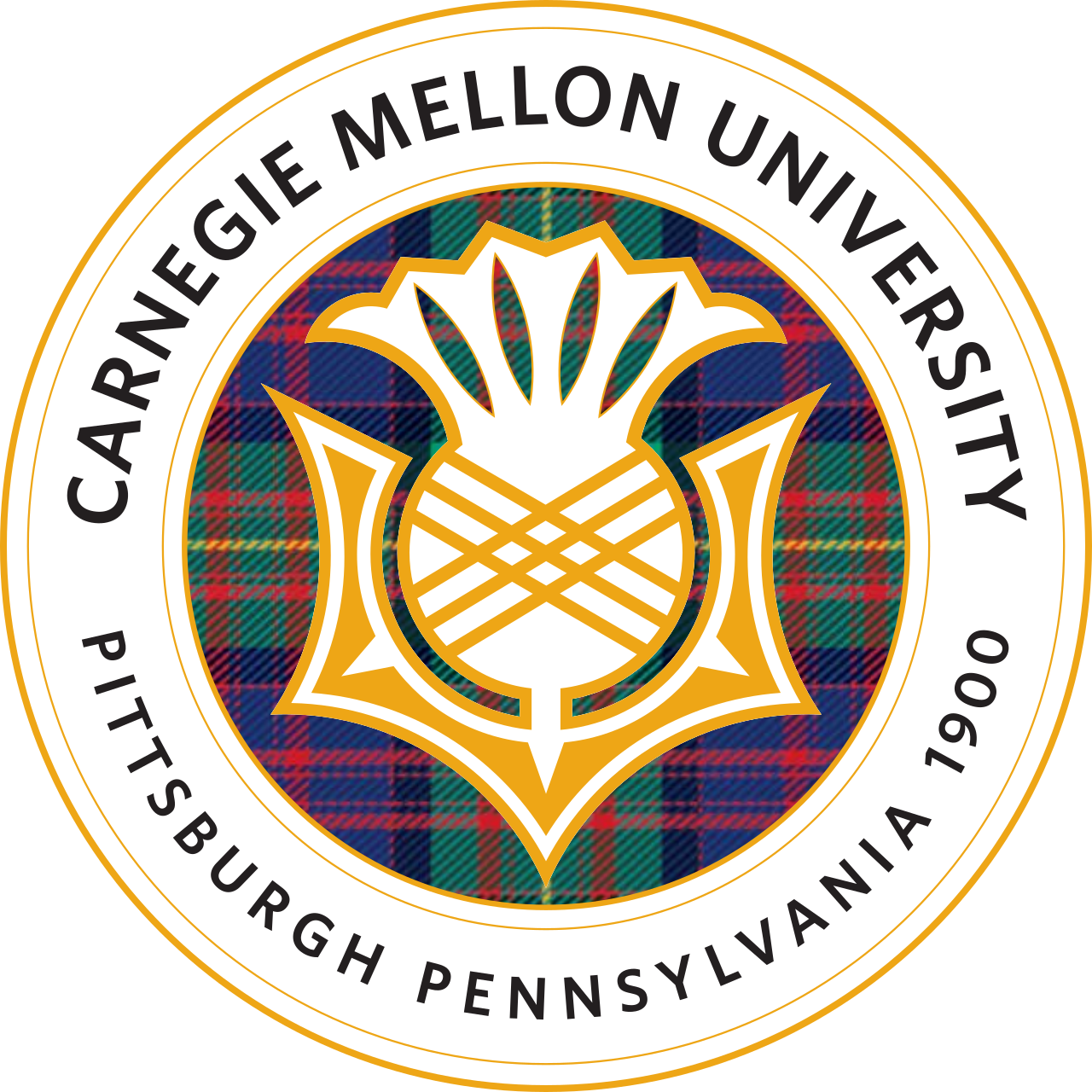}
    \end{minipage}%
    \hspace{-1cm}
    \begin{minipage}[c]{0.5\linewidth}
        \centering
        \footnotesize\textcolor{gray!70}{Preprint \\ \today }
    \end{minipage}%
    \begin{minipage}[c]{0.25\linewidth}
        \hfill 
    \end{minipage}
    
    \vspace{0.25cm}
    {\color{DeepIndigo!20}\rule{\linewidth}{0.6pt}}
    
    \vspace{0.4cm}
\end{center}

\begin{center}
{\LARGE \bfseries
QueenBee: A Recursively Self-Improving Harness for Communication Topology Design in LLM-Based Multi-Agent Systems\\
\par}

\vspace{0.25cm}

{\large
Congjia Tian$^{1}$ \quad
Yuhang Yao$^{2}$ \quad
Jiaming Cui$^{1}$ \\}
\vspace{0.05cm}
{$^{1}$\texttt{Virginia Tech, Blacksburg, VA}\\}
{$^{2}$\texttt{Carnegie Mellon University, Pittsburgh, PA}\\}
{\small
\texttt{\{robintian777, jiamingcui\}@vt.edu, yuhangyao8@gmail.com}\\}

\vspace{0.15cm}

{\small \faGithub \quad \href{https://github.com/RobinTian-7/QueenBeePlanner}{\textcolor{academicblue}{\texttt{https://github.com/RobinTian-7/QueenBeePlanner}}}}

\vspace{0.5cm}
\end{center}

\renewenvironment{abstract}{\noindent\ignorespaces}{\par}

\begin{center}
\begin{tcolorbox}[
    enhanced,
    colback=caspianbg,
    colframe=gray!5,
    arc=0mm,
    outer arc=0mm,
    width=\linewidth,
    left=4mm,
    right=4mm,
    top=3mm,
    bottom=3mm,
    boxrule=0pt,
    borderline west={2.5pt}{0pt}{DeepIndigo!90!ElectricCyan}, 
    before skip=10pt,
    after skip=20pt
]
LLM-based multi-agent systems show great potential for complex tasks through role specialization and collaborative reasoning. However, existing automated workflow design methods primarily focus on one-off search and candidate optimization, lacking sustained accumulation and reuse of structural design experience to meet evolving coordination demands. We propose QueenBee, an experience-guided framework for multi-agent workflow evolution. QueenBee represents workflows as executable programs that generate temporal communication DAGs and maintains Structural Feedback and Transition Bank (SFT-Bank), a persistent structural memory of execution diagnoses, design hypotheses and validation outcomes, and historical programs. This accumulated experience guides subsequent workflow generation, validation, and optimization, enabling the co-evolution of collaboration programs and design experience. Experiments on Silo-Bench show that QueenBee outperforms fixed communication baselines across team sizes and achieves the highest mean task score among recent automated multi-agent workflow optimization methods. Ablation studies further show that SFT-Bank components provide complementary performance gains, while evolution dynamics reveal a consistently improving average trend in the quality of the best collaboration structures found across independent runs as structural experience accumulates.
\end{tcolorbox}
\end{center}

\section{Introduction}

LLM-based agents have shown strong abilities in reasoning, planning, and coding \citep{guo2025deepseekr1,yao2023react,yang2024sweagent}. Yet real-world tasks often involve interdependent subtasks, heterogeneous tools, feedback from dynamic environments, and long-term information management \citep{schick2023toolformer,hong2024metagpt,wu2024autogen}. Single agents still struggle with reliable long-term coordination and systematic verification \citep{liu2024agentbench,liu2024lost}. Multi-agent systems (MAS) offer an effective approach to complex tasks through role specialization, collaboration, and information exchange \citep{hong2024metagpt,wu2024autogen,li2023camel,du2024improving,wang2025mixtureofagents}.

Despite their success in collaborative reasoning and complex tasks, MAS performance depends heavily on workflow design. Existing frameworks often require experts to specify roles, communication topologies, execution order, and work instructions \citep{hong2024metagpt,wu2024autogen,li2023camel}. However, fixed structures adapt poorly to different tasks and team sizes, while poor communication may cause information loss, redundant computation, and inconsistent conclusions \citep{zhang2026silobench}. Manual design is therefore costly and poorly suited to changing collaboration needs.

Automated MAS optimization methods mainly fall into three categories. \qbemph{Topology optimization methods}, including GPTSwarm, G-Designer, MaAS, and MASS, optimize communication links and system architectures \citep{zhuge2024gptswarm,zhang2025gdesigner,zhang2025multiagent,zhou2026multiagent} but lack direct program-level modeling of information transfer and integration across stages. \qbemph{Code and workflow search methods}, including ADAS, AFlow, and A2Flow, expand the design space through program generation, execution feedback, and workflow search \citep{hu2025automated,zhang2025aflow,zhao2026a2flow} but do not specifically accumulate structural modification experience from MAS collaboration failures. \qbemph{Evolutionary search methods}, such as EvoMAS, improve MAS configurations through mutation, crossover, and experience memory \citep{hu2026evomas} but have yet to organize temporal collaboration outcomes into unified, traceable design knowledge.

Prior work on experiential learning and agent self-evolution shows that systems can improve through external memory, reusable skills, and program modifications without updating model parameters \citep{shinn2023reflexion,zhao2024expel,wang2024voyager,wang2025agent,zhang2026darwin,lee2026metaharness,lin2026agentic}. However, \qbemph{MAS optimization requires not only better configurations but also learning to design better coordination programs.} We argue that past execution results should become knowledge that directly guides future structural design,  allowing \qbemph{coordination programs and design experience to co-evolve}.

\begin{figure*}[t]
\renewcommand{\thefigure}{1}
\centering
\includegraphics[width=\textwidth]{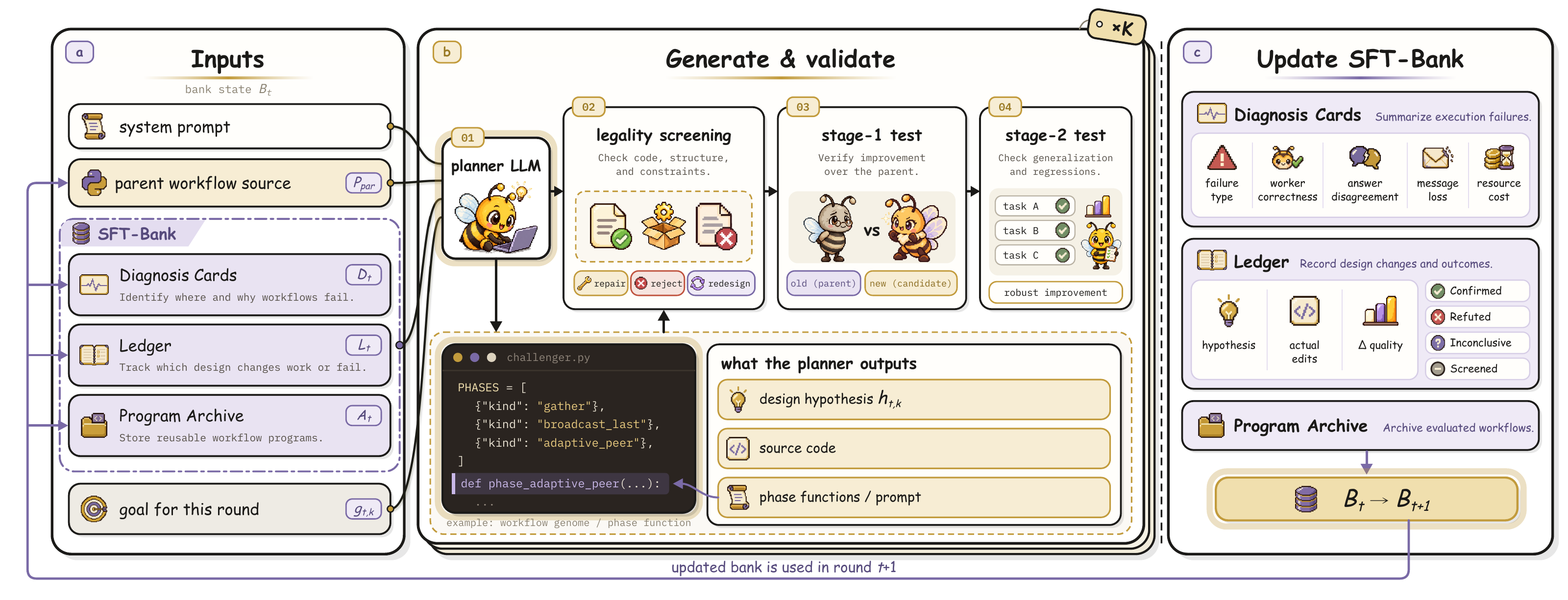}%
\caption{\textbf{Overview of QueenBee's experience-guided workflow evolution.} The Planner receives the parent workflow, exploration goal, and experience retrieved from SFT-Bank. It generates a design hypothesis and workflow code, followed by legality screening, parent-comparison testing, and cross-task validation. Execution feedback updates Diagnosis Cards, the Ledger, and the Program Archive, guiding subsequent evolution.}
\end{figure*}

We therefore represent MAS workflows as executable programs that generate \qbemph{temporal communication DAGs}, enabling joint optimization and inheritance of communication structures and work instructions. Based on this representation, we propose \qbemph{QueenBee}, an experience-guided workflow evolution framework that uses persistent structural memory, \qbemph{Structural Feedback and Transition Bank (SFT-Bank)}, to accumulate execution traces, design hypotheses and their validation results, and historical programs, enabling workflows to benefit continuously from past experience.

We evaluate QueenBee on Silo-Bench and other problems across team sizes, model configurations, and automated optimization methods, and examine its experience components through ablation studies. Results show that QueenBee improves collaboration quality across team sizes and achieves the highest overall task performance among the compared automated workflow optimization methods.

Our main contributions are: (1) \qbemph{Executable Workflow Evolution Paradigm.} We represent MAS workflows as programs generating temporal communication DAGs, enabling joint optimization of communication structures and work instructions. (2) \qbemph{QueenBee and SFT-Bank.} We propose an experience-guided workflow evolution framework that continuously improves workflows by accumulating and reusing design experience through structured memory. (3) \qbemph{Systematic Experimental Evaluation.} We demonstrate QueenBee's effectiveness across team sizes and model configurations on Silo-Bench and assess component contributions through ablation studies.

\section{Related Work}

\subsection{Multi-Agent Communication Topology}

Early LLM-based MAS relied on handcrafted collaboration: CAMEL uses role-playing \citep{li2023camel}, MetaGPT assigns specialized development roles \citep{hong2024metagpt}, AutoGen enables programmable agent conversations \citep{wu2024autogen}, Multi-Agent Debate uses multi-round discussion \citep{du2024improving}, and Mixture-of-Agents employs layered aggregation \citep{wang2025mixtureofagents}.

Later methods automate topology design. GPTSwarm jointly optimizes prompts and connections \citep{zhuge2024gptswarm}, DyLAN dynamically selects agents \citep{liu2024dynamic}, AgentPrune removes redundant edges \citep{zhang2025cut}, G-Designer generates task-conditioned topologies \citep{zhang2025gdesigner}, MaAS searches architectures \citep{zhang2025multiagent}, and MASS jointly optimizes prompts and topologies \citep{zhou2026multiagent}. However, these methods primarily optimize agent selection, connections, or architectures, without explicitly modeling cross-round information propagation and integration as executable coordination programs. Unlike these methods, QueenBee encodes communication timing, information processing, and work instructions as executable coordination programs.

\subsection{Automated Workflow Optimization}

ADAS generates code-based agents and archives designs \citep{hu2025automated}, AFlow combines Monte Carlo Tree Search with code modification and execution feedback \citep{zhang2025aflow}, AgentSquare searches modular agent configurations \citep{shang2025agentsquare}, and A2Flow introduces reusable abstractions \citep{zhao2026a2flow}. EvoMAS evolves declarative MAS configurations through trace-guided mutation, crossover, candidate pools, and experience memory \citep{hu2026evomas}. MAS-Architect separates topology planning from node implementation \citep{huang2026masarchitect}, while DAAO adapts workflow depth, operators, and model routing \citep{su2026difficultyaware}. However, these approaches lack a unified mechanism for organizing structural modifications and their validation outcomes into reusable experience and memory.

Unlike these methods, QueenBee evolves \qbemph{inheritable temporal coordination programs}, linking failure diagnoses, structural modifications, and experimental validation into reusable design knowledge.

\subsection{Experience-Based Self-Improvement and Harness Evolution}

External memory and skills enable improvement without parameter updates. Reflexion stores verbal feedback \citep{shinn2023reflexion}, ExpeL extracts execution experience \citep{zhao2024expel}, Voyager builds executable skill libraries \citep{wang2024voyager}, and Agent Workflow Memory retrieves historical workflows \citep{wang2025agent}. Promptbreeder evolves prompts \citep{fernando2024promptbreeder}, STOP modifies optimization scaffolds \citep{zelikman2024selftaught}, GEPA uses trajectory reflection \citep{agrawal2026gepa}, Agentic Context Engineering evolves contextual experience \citep{zhang2026agentic}, and Meta Context Engineering co-evolves context skills and artifacts \citep{ye2026meta}.

Program and harness evolution extend this approach. Darwin Gödel Machine evolves agent code \citep{zhang2026darwin}, AlphaEvolve combines code generation and evolutionary selection \citep{novikov2025alphaevolve}, Meta-Harness optimizes execution frameworks \citep{lee2026metaharness}, and Self-Harness validates modifications through failure diagnosis and regression testing \citep{zhang2026selfharness}. Agentic Harness Engineering emphasizes observability and reversibility \citep{lin2026agentic}, Continual Harness adapts execution components online \citep{karten2026continual}, and DemoEvolve uses demonstrations under sparse feedback \citep{che2026demoevolve}. Harness Updating Is Not Harness Benefit distinguishes updating capability from actual gains \citep{lin2026harness}. However, these approaches primarily improve agent capabilities or general execution frameworks and QueenBee applies experience-driven improvement to MAS coordination through \qbemph{execution diagnoses, validated design hypotheses, and inheritable programs}.

\section{Method}

\subsection{Problem Formulation}

\qblead{Multi-Agent Workflow.} Consider $n$ LLM-based Workers $\mathcal{W}=\{W_1,\ldots,W_n\}$ jointly solving tasks sampled from $\mathcal{D}$. Each Worker $W_i$ observes only $o_i$, its private task information, and exchanges information over discrete communication rounds indexed by $\tau$. With fixed Worker parameters, an executable Python workflow $P$ determines $(a_{i,\tau},R_{i,\tau},\omega_{i,\tau})=P(i,\tau,s_{i,\tau})$, where $a_{i,\tau}$ specifies whether to send, $R_{i,\tau}\subseteq\mathcal{W}\setminus\{W_i\}$ specifies recipients, $\omega_{i,\tau}$ gives optional semantic instructions, and $s_{i,\tau}$ is the observable state, including the round, known information sources, and pending messages. The program also determines the termination round $T_{\mathrm{comm}}$, after which every Worker submits its own final answer.

\qblead{Temporal Communication DAG.} Executing $P$ on task $x$ produces a time-unrolled DAG:
\qbdisplay{%
G_{P,x}=\operatorname{Unroll}(P,x)=(V,E_{\mathrm{comm}}\cup E_{\mathrm{state}}),
}
where $V=\{(i,\tau)\mid i\in[n],\ 0\leq\tau\leq T_{\mathrm{comm}}\}$. Each node $(i,\tau)$ represents Worker $W_i$ at round $\tau$. Communication edges link a sender $(i,\tau)$ to next-round recipients $(j,\tau+1)$; state edges connect consecutive states of the same Worker. All edges advance one round, ensuring acyclicity. Topology controls information flow and timing, while semantic instructions control its processing and expression.

\qblead{Executable Representation and Objective.} The executable program has two policy functions: $f_{\mathrm{comm}}$ selects communication actions, recipients, and semantic instructions from the round, Worker identity, and observable state; $f_{\mathrm{stop}}$ determines termination. Its editable Genome contains communication phases, routing policies, and semantic work instructions. The executor unrolls these policies from task inputs and runtime states, while Worker parameters, the underlying communication protocol, and execution environment remain fixed. We optimize executable programs through inheritance, modification, and combination. QueenBee repeatedly generates, executes, and evaluates candidates to accumulate reusable experience that improves subsequent search. The design model remains fixed; only its external experience changes. The formal design-capability objective is given in Appendix E.

\subsection{SFT-Bank as Structural Memory}

We introduce the Structural Feedback and Transition Bank (SFT-Bank) to provide persistent structural memory for workflow design. SFT-Bank preserves execution diagnoses, design hypotheses and their validation outcomes, and historical executable programs. Formally, at iteration $t$, the bank consists of three components, denoted as $\mathcal{B}_t=(\mathcal{D}_t,\mathcal{L}_t,\mathcal{A}_t)$.

\qblead{Diagnosis Cards $\mathcal{D}_t$.} Diagnosis Cards record task performance, Worker correctness distributions, answer disagreements, information loss, and stage-level resource use, organized by propagation, aggregation, and final-output failures.

\qblead{Ledger $\mathcal{L}_t$.} The Ledger records each design hypothesis and validation outcome, linking parent and candidate workflows, target tasks, intended improvements, and observed performance changes. Attempts are labeled Confirmed, Refuted, or Inconclusive based on empirical comparisons; screened proposals are also retained.

\qblead{Program Archive $\mathcal{A}_t$.} The Program Archive stores executable source code and metadata for strong overall programs, task-specific specialists, and successful or failed implementations.

\subsection{Experience-Guided Workflow Evolution}

Each iteration uses SFT-Bank to schedule exploration and retrieve historical experience. The fixed-parameter Planner generates candidates, which the harness evaluates and uses to update SFT-Bank.

\qblead{Goal Scheduling.} For each of $K$ proposals, the harness selects a goal $g_{t,k}$ and parent workflow $P_{t,k}^{\mathrm{par}}$ from $\mathcal{B}_t$. Most proposals address unresolved training-task failures grouped by output format and dominant failure type. An Upper Confidence Bound (UCB) strategy prioritizes groups by remaining improvement potential, historical quality changes, and attempt counts. Repeatedly refuted directions receive lower priority, and each iteration favors distinct groups. The harness selects strong task-specific parents from the Program Archive $\mathcal{A}_t$ so candidates inherit different local strengths. Other slots periodically explore Merge, combining complementary workflows across tasks; Economize, reducing model calls and tokens while preserving quality; and Explore, seeking alternative designs for unresolved, underexplored tasks.

\qblead{Experience-Guided Proposal.} For candidate $k$, the harness constructs a design context from the three SFT-Bank components:
\qbdisplay{%
\begin{aligned}
\mathcal{C}_{t,k}
&=\operatorname{Retrieve}(\mathcal{B}_t,g_{t,k},P_{t,k}^{\mathrm{par}})\\
&=(\mathcal{C}_{t,k}^{D},\mathcal{C}_{t,k}^{L},\mathcal{C}_{t,k}^{A}).
\end{aligned}
}

Diagnosis Cards $\mathcal{D}_t$ provide the parent's task-specific failure patterns; the Ledger $\mathcal{L}_t$ supplies recently supported hypotheses, repeatedly ineffective modifications, and their measured effects; and the Program Archive $\mathcal{A}_t$ supplies successful and failed implementations addressing similar failures, plus complementary designs for workflow combinations. Retrieved programs are presented as editable code examples. The Planner also receives the parent's editable Python Genome, its training-task quality and cost records, the exploration goal, and fixed interface constraints $\mathcal{I}$. It produces a design hypothesis $h_{t,k}$ specifying the target problem, modification mechanism, and expected improvement, together with a new Python Genome $\gamma_{t,k}$:
\qbdisplay{%
(h_{t,k},\gamma_{t,k})=\operatorname{Planner}_{\theta}
(g_{t,k},\gamma_{t,k}^{\mathrm{par}},\mathcal{C}_{t,k},\mathcal{I}).
}

The harness embeds the Genome into the fixed framework to form candidate $P_{t,k}$, which generates a temporal communication DAG from task inputs and runtime states. Planner and Worker weights remain fixed; only workflow programs change. Retrieval and scheduling details appear in Appendix E.

\subsection{Empirical Evaluation and Experience Consolidation}

\qblead{Staged Empirical Evaluation.} The harness first screens code and structural validity, rejecting duplicates, constraint violations, and historically ineffective modifications. \textbf{In Stage 1}, it executes each screened candidate on target training tasks and reruns its parent under identical conditions; only candidates meeting predefined quality requirements without output-format errors advance. \textbf{In Stage 2}, additional training tasks spanning diverse templates and output formats test cross-task effectiveness, while regressions on previously strong tasks are checked. New task-specific specialists may receive independent confirmation tests to reduce chance effects. Evaluation records task quality, execution cost, parent-relative changes, and task-level gains and regressions.

\qblead{Experience Consolidation.} Updates derive from execution evidence and source-code differences, not Planner self-assessment. Diagnosis Cards $\mathcal{D}_t$ receive execution statistics grouped by failure type, excluding specific answers and ground-truth labels for safe reuse. By comparing parent and candidate source code, the harness identifies added or removed communication stages, changes to round scheduling, and revised work instructions. The Ledger $\mathcal{L}_t$ associates these changes with task-level performance and regressions, labeling hypotheses Confirmed, Refuted, or Inconclusive; unexecuted proposals are Screened. Confirmation requires a predefined improvement margin without excessive regressions, while substantial target-task degradation yields refutation. The Ledger also tracks repeatedly refuted modification directions. The Program Archive retains programs, parent-version relationships, training evaluations, and metadata, preserving task-specific strengths for later reuse. With feedback $e_{t,k}$ for each candidate, the update is:
\qbdisplay{%
\mathcal{B}_{t+1}=U\left(\mathcal{B}_t,\{(h_{t,k},P_{t,k},e_{t,k})\}_{k=1}^{K}\right).
}

Screened and inconclusive attempts may still leave records; SFT-Bank updates do not require a candidate to outperform existing workflows. The resulting memory guides both the Planner's design context and future goal and parent selection. Validation criteria and consolidation rules are detailed in Appendix E.

\section{Experiment}

\subsection{Experimental Setup}

We primarily evaluate QueenBee on Silo-Bench and further assess the reliability of information propagation and aggregation using the Count-Frequency distributed statistics task (Appendix).

\qblead{Benchmarks and Tasks.} Silo-Bench distributes task information across Workers, requiring them to integrate private information through communication to solve tasks beyond any individual Worker's capability. It comprises 30 task types, three difficulty levels, and varying computational requirements. We use 18 development task types and 12 independent test task types.

\qblead{Baselines.} Fixed communication baselines include Silo-Bench's round-robin peer-to-peer communication (P2P), Broadcast, and staged forwarding (SFS), alongside the unevolved initial workflow $P_0$. Automated optimization baselines include GPTSwarm \citep{zhuge2024gptswarm}, G-Designer \citep{zhang2025gdesigner}, MaAS \citep{zhang2025multiagent}, and EvoMAS \citep{hu2026evomas}. We also assess the role of historical experience through memory-free search and SFT-Bank component ablations.

\qblead{Model and Implementation Settings.} We use GPT-5.1 and GPT-5 mini \citep{openai2025gpt5}, DeepSeek-V4.1-Flash \citep{deepseekai2026deepseekv4}, and gpt-oss-120b \citep{openai2025gptoss120b} models as Planners or Workers across team sizes and model combinations. Controlled comparisons vary only the target variable, and Worker model parameters remain fixed throughout evolution. Model versions, inference settings, and search budgets are specified in the corresponding experiments.

\qblead{Experimental Protocol.} Development data support candidate generation, execution evaluation, and SFT-Bank updates. Independent validation data, when used, serve only for post-evolution workflow selection. Test data are excluded from search, experience updates, and selection, and are used only for independent evaluation. All comparisons use identical test tasks, with budgets and repetition counts specified in the corresponding sections.

\qblead{Evaluation Metrics.} We measure partial completion score (PCS). PCS is the benchmark's partial-credit score: each Worker's answer receives a task-specific quality score in [0, 1], averaged over Workers and test executions. Execution and search costs are measured by per-task and total search token consumption, respectively.

\subsection{Main Result}

\begin{table*}[t]
\centering
\begingroup\small
\setlength{\tabcolsep}{8pt}
\begin{adjustbox}{max width=\textwidth}
\begin{tabular}{lcccccc}
\toprule
 & \multicolumn{3}{c}{gpt-oss-120b} & \multicolumn{3}{c}{DeepSeek-V4.1-Flash} \\
\cmidrule(lr){2-4}\cmidrule(lr){5-7}
Method & $n=2$ & $n=5$ & $n=10$ & $n=2$ & $n=5$ & $n=10$ \\
\midrule
\multicolumn{7}{l}{\textit{Fixed Protocols}} \\
\quad P2P                &  29.2 &  27.5 &  10.0 &  69.0 &  71.3 &  59.6 \\
\quad Broadcast          &  47.9 &  40.8 &  23.8 &  79.2 &  72.9 &  70.8 \\
\quad SFS                &  12.5 &   5.0 &   4.6 &  75.6 &  72.4 &  72.8 \\
\quad Initial Workflow   &  41.7 &  17.5 &   5.4 &  66.7 &  64.1 &  57.2 \\
\multicolumn{7}{l}{\textit{Automated Workflow Optimization}} \\
\quad Random Search      &  58.3 &  25.0 &  20.6 &  70.8 &  78.3 &  71.7 \\
\quad GPTSwarm           &  52.8 &  28.3 &  28.1 &  71.8 &  75.6 &  60.8 \\
\quad G-Designer         &  40.3 &  21.7 &  16.9 &  57.6 &  78.3 &  75.8 \\
\quad EvoMAS             &  77.8 &  42.8 &  46.9 &  79.2 &  73.5 &  66.7 \\
\quad MaAS               &  13.9 &  13.3 &  23.9 &  68.1 &  61.1 &  82.5 \\
\textbf{QueenBee (Ours)} & \textbf{79.2} & \textbf{54.4} & \textbf{51.7} & \textbf{80.6} & \textbf{83.3} & \textbf{83.7} \\
\bottomrule
\end{tabular}
\end{adjustbox}
\endgroup
\renewcommand{\thetable}{1}
\caption{QueenBee versus fixed communication protocols and automated workflow optimization methods with 2, 5 and 10 Workers. Entries report the GPT-OSS-120b configuration (as both  Planner and Workers) and the DeepSeek configuration (as both  Planner and Workers), respectively; fixed protocols use the same Workers. Results under the GPT configuration (GPT-5.1 Planner, GPT-5 mini Workers) are reported in Table H1.}
\end{table*}

We evaluate QueenBee against fixed communication protocols and automated workflow optimization methods on Silo-Bench with 2, 5, and 10 Workers, with results reported in Table 1.

\qblead{QueenBee outperforms fixed communication protocols.} We compare QueenBee with P2P, Broadcast, SFS, and the initial workflow under gpt-oss-120b and DeepSeek-V4.1-Flash Planner / Worker configurations. With 5 Workers, QueenBee achieves 54.4\% SR under gpt-oss and 83.3\% PCS under DeepSeek, surpassing the best fixed protocol, Broadcast, by 13.6 and 10.4 percentage points, respectively. QueenBee achieves the highest task quality in both configurations, showing that jointly optimizing communication structures and information processing improves collaboration within fixed teams.

\qblead{QueenBee outperforms existing automated workflow optimization methods.} We further compare QueenBee with Random Search, GPTSwarm, G-Designer, EvoMAS, and MaAS using gpt-oss-120b and DeepSeek-V4.1-Flash, identical task splits, and a budget of 80 effective task executions. With 5 Workers, QueenBee achieves the highest scores under both models (54.4\% SR under gpt-oss and 83.3\% PCS under DeepSeek), exceeding the strongest baselines, EvoMAS (42.8\%) and Random Search and G-Designer (both 78.3\%), by 11.6 and 5.0 percentage points, respectively. 

Overall, QueenBee outperforms fixed protocols across model configurations and achieves the highest overall task performance among automated optimization methods, demonstrating the effectiveness of experience-guided executable workflow evolution.

\subsection{Sensitivity Analysis}

\subsubsection{Sensitivity to Team Size}

Having established QueenBee's advantages over existing baselines with 5 Workers, we examine whether its improvements persist across team sizes and model configurations.

\qblead{QueenBee improves task quality across team sizes.} As Table 2 shows, we evaluate teams of 2, 5, and 10 Workers under GPT-5.1 / GPT-5 mini and gpt-oss-120b configurations. Under GPT, QueenBee improves PCS from 58.3\%, 51.4\%, and 8.6\% to 70.8\%, 58.5\%, and 58.5\%, respectively; under gpt-oss, from 50.0\%, 40.0\%, and 34.3\% to 64.3\%, 60.0\%, and 60.0\%. Consistent gains across all settings demonstrate stable optimization across team sizes and models.

\begin{table}[h]
\centering
\begingroup\small
\setlength{\tabcolsep}{21pt}
\begin{adjustbox}{max width=\columnwidth}
\begin{tabular}{@{}lr@{\,$\rightarrow$\,}lr@{\,$\rightarrow$\,}l@{}}
\toprule
Team size & \multicolumn{2}{c}{\makebox[0pt]{GPT-5.1 / GPT-5 mini}} & \multicolumn{2}{c}{\makebox[0pt]{gpt-oss-120b}} \\
\midrule
N=2 & 58.3 & 70.8 & 50.0 & 64.3 \\
N=5 & 51.4 & 58.5 & 40.0 & 60.0 \\
N=10 & 8.6 & 58.5 & 34.3 & 60.0 \\
\bottomrule
\end{tabular}
\end{adjustbox}
\endgroup
\renewcommand{\thetable}{2}
\caption{Test PCS (\%) of the initial and final workflows (initial → final) across team sizes and models.}
\end{table}

\qblead{Larger teams benefit more from structural optimization.} Initial PCS declines markedly with team size under both model configurations, whereas QueenBee mitigates this degradation. For 10-Worker teams, PCS increases by 49.9 and 25.7 percentage points under GPT and gpt-oss, respectively, the largest gains within each configuration. These results indicate that experience-guided structural optimization yields greater quality improvements as team size increases, consistently across both models.

\subsubsection{Sensitivity to Model Configurations}

We examine QueenBee's sensitivity to model configurations by independently evolving workflows with DeepSeek-V4.1-Flash-thinking-max and gpt-oss-120b Planners, each paired with Workers of the same model family. Both use 5 Workers and a budget of 80 effective task executions. The resulting workflows are frozen and evaluated with both Worker models on 12 held-out task types.

\qblead{QueenBee achieves consistent gains across model configurations.} As Table 3 shows, all four Planner–Worker combinations outperform their initial workflows. Notably, workflows evolved with the gpt-oss Planner achieve 75.0\% PCS with DeepSeek Workers and 61.2\% with gpt-oss Workers, compared with initial scores of 66.1\% and 19.9\%. These results demonstrate that QueenBee remains effective across model configurations and that evolved workflows can transfer to different Worker models.

\begin{table}[h]
\centering
\begingroup\small
\setlength{\tabcolsep}{7pt}
\begin{adjustbox}{max width=\columnwidth}
\begin{tabular}{@{}llr@{\,$\rightarrow$\,}l@{}}
\toprule
Planner & Worker & \multicolumn{2}{c}{\makebox[0pt]{PCS (\%)}} \\
\midrule
DeepSeek & DeepSeek & 66.1 & 83.1 \\
DeepSeek & gpt-oss & 19.9 & 66.1 \\
gpt-oss & DeepSeek & 66.1 & 75.0 \\
gpt-oss & gpt-oss & 19.9 & 61.2 \\
\bottomrule
\end{tabular}
\end{adjustbox}
\endgroup
\renewcommand{\thetable}{3}
\caption{Test PCS (\%) of initial and evolved workflows (initial → final) across Planner–Worker configurations. Workflows are evolved with matched Planner–Worker model families and evaluated using either Worker model.}
\end{table}

\qblead{Performance is sensitive to model configurations.} With the test Worker fixed, switching from DeepSeek-evolved to gpt-oss-evolved workflows reduces PCS by 8.1 and 4.9 percentage points for DeepSeek and gpt-oss Workers, respectively. Switching the test Worker while keeping the evolved workflow fixed produces larger differences of 17.0 and 13.8 points. Thus, execution-model choice has a greater observed impact than the evolution configuration in this experiment.

\qblead{Planner capability affects design precision.} The DeepSeek Planner replaces the initial relay chain with centralized gathering and broadcasting, explicitly preserving data sources, order, and raw content so Workers independently compute from a consistent information snapshot. By contrast, the gpt-oss Planner retains the relay structure, adding a digest phase and generic forwarding instructions without explicit data-preservation constraints. In one III-29 test run, incomplete information propagation caused all five Workers to submit distinct incorrect answers. This case illustrates the importance of translating historical diagnoses into precise executable instructions, beyond modifying communication topology alone. Program details and execution traces are provided in the appendix.

\subsection{Performance over Evolution}

To assess QueenBee's ability to continually improve collaboration structures, we examine workflow performance across evolution rounds for teams of 2, 5, and 10 Workers. As Figure 2 shows, historical workflows are evaluated retrospectively on held-out tasks after evolution, with test data excluded from candidate generation, validation, experience updates, and selection. The trajectories thus reflect how discovered workflows perform on unseen tasks.

\begin{figure}[t]
\renewcommand{\thefigure}{2}
\centering
\includegraphics[width=0.5\columnwidth]{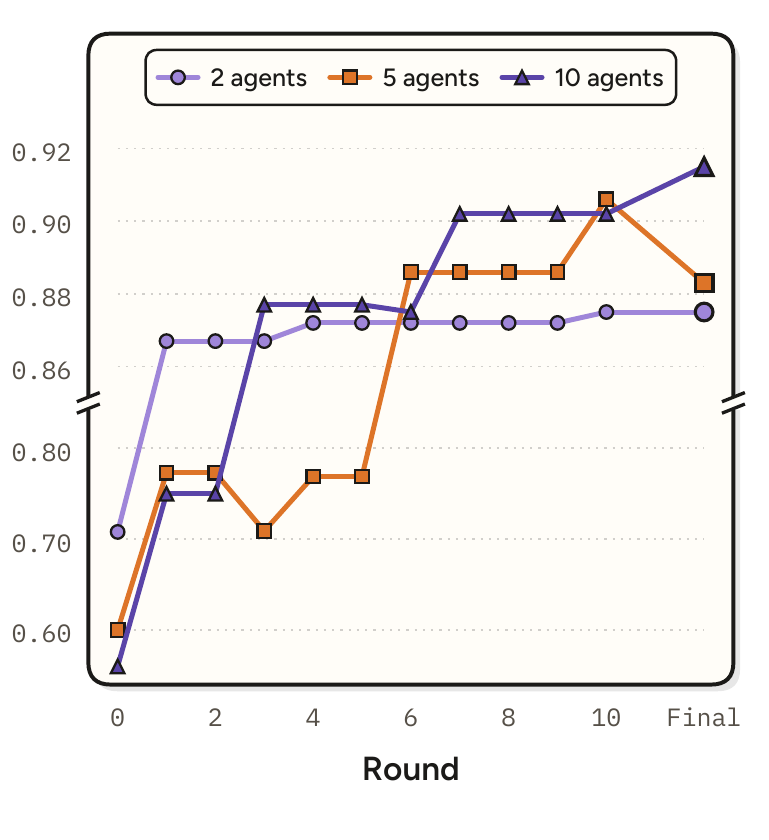}%
\caption{Evolution dynamics of QueenBee with DeepSeek-V4.1-Flash as Planner and Workers. Held-out test PCS of the best workflow at each evolution round and of the final selected workflow, for teams of 2, 5, and 10 Workers.}
\end{figure}

\qblead{Structural evolution yields sustained performance gains.} As Figure 2 shows, mean PCS rises stepwise across all three team sizes. Smaller teams saturate early, whereas larger teams continue to improve in later rounds. Some rounds yield no improvement, while effective structures raise performance to new plateaus. Through candidate generation, staged validation, and historical program retention, QueenBee converts continued exploration into cumulative workflow improvements.

\qblead{Larger teams gain more from evolution.} With more Workers, each Worker handles a smaller shard of the task and completes it more reliably, but the initial propagation structure integrates these distributed results poorly, so initial PCS declines as team size increases. As evolution proceeds, this propagation disadvantage is gradually worn away: the Planner learns from accumulated experience how structural changes affect performance and combines effective structures, so that trial and error becomes directed. The advantage of better per-shard execution then emerges. The 10-Worker team, which starts from the lowest initial performance, achieves the largest gain and the highest final PCS, reversing the initial ordering across team sizes.

\qblead{Evolution gains stem from coordination optimization.} Each evolution run fixes the Worker model and task configuration, optimizing only communication structures and work instructions. Thus, observed gains reflect improved workflow organization rather than changes in underlying model capabilities. The improving trajectories and the reversed ordering across team sizes indicate that QueenBee progressively discovers and retains higher-quality executable collaboration workflows through repeated exploration, empirical validation, and reuse of historical designs.

\subsection{Ablation Study}

To assess SFT-Bank's component contributions, we separately remove Diagnosis Cards $\mathcal{D}_t$, the Ledger $\mathcal{L}_t$, and the Program Archive $\mathcal{A}_t$, alongside a memory-free variant (w/o SFT-Bank). The latter retains candidate generation, execution evaluation, and elite selection but excludes experience accumulated across iterations.

All ablations use the setting of Section 4.2.

\qblead{SFT-Bank improves workflow optimization.} As Table 4 shows, full QueenBee achieves the highest SR (80.6\%) and PCS (83.1\%). Removing SFT-Bank reduces these to 63.9\% and 66.7\%, drops of 16.7 and 16.4 percentage points. Thus, structured experience accumulated across iterations improves final workflow quality beyond basic search and selection, providing reusable history for subsequent structural design.

\begin{table}[h]
\centering
\begingroup\small
\begin{adjustbox}{max width=\columnwidth}
\begin{tabular}{lrrr}
\toprule
Method & SR (\%) ↑ & PCS (\%) ↑ & ΔPCS \\
\midrule
\textbf{QueenBee (Full)} & \textbf{80.6} & \textbf{83.1} & — \\
w/o Diagnosis & 77.8 & 77.8 & −5.3 \\
w/o Ledger & 75.0 & 76.1 & −7.0 \\
w/o Archive & 63.9 & 65.4 & −17.7 \\
w/o SFT-Bank & 63.9 & 66.7 & −16.4 \\
\bottomrule
\end{tabular}
\end{adjustbox}
\endgroup
\renewcommand{\thetable}{4}
\caption{Ablation results for SFT-Bank components.}
\end{table}

\qblead{All experience components contribute to performance.} Removing Diagnosis Cards, the Ledger, and the Program Archive reduces PCS by 5.3, 7.0, and 17.7 percentage points, respectively. Without Diagnosis Cards $\mathcal{D}_t$, the Planner cannot use historical failure types and diagnoses to guide structural improvements; without the Ledger $\mathcal{L}_t$, it loses systematic records and use of past structural modifications and validation results. Both reduce task quality, indicating the value of diagnosing and organizing execution history.

\qblead{Removing the Program Archive causes the largest drop.} Among single-component ablations, w/o Archive performs worst (63.9\% SR, 65.4\% PCS). It removes both historical program examples and archive-based parent selection, concentrating search on the current incumbent. This suggests that retaining and reusing historical programs is important for experience-guided search: inheritance, combination, and modification let the Planner exploit past designs beyond a single incumbent. Its score slightly below w/o SFT-Bank reflects run-to-run LLM fluctuation, not a negative contribution.

Overall, full SFT-Bank outperforms all ablations. The Program Archive $\mathcal{A}_t$ and associated structural inheritance have the greatest impact, while Diagnosis Cards and the Ledger provide additional gains. Representative workflow analysis also reveals distinct information-processing strategies across experience configurations.

\section{Conclusion}

We propose \qbemph{QueenBee}, an experience-guided multi-agent workflow evolution framework that represents workflows as executable programs generating temporal communication DAGs. Through persistent structural memory, \qbemph{SFT-Bank}, QueenBee continuously accumulates execution traces, design hypotheses and their validation results, and historical programs, enabling workflow design to reuse past experience. This transforms execution results into knowledge guiding subsequent structural design, allowing \qbemph{coordination programs and design experience to co-evolve}. Experiments on Silo-Bench show that QueenBee improves collaboration quality across team sizes and achieves the highest overall task performance among the compared automated workflow optimization methods.

\section*{Limitations}

This study has several limitations. First, QueenBee is evaluated mainly on controlled tasks such as Silo-Bench and Count-Frequency; its generalization to open-ended real-world tasks and dynamic environments remains unverified. Second, workflow evolution requires repeated candidate generation and empirical evaluation, incurring additional search costs, while performance gains in some configurations entail higher execution costs. Furthermore, the experiments cover limited task scales, repetitions, and model combinations, providing insufficient evidence of experience transfer across tasks and models and no guarantee of sustained performance gains throughout evolution. Future work will explore broader applications, cross-task experience reuse, and more efficient evolution strategies .Finally, repeated LLM-based search may increase energy consumption, while deploying automatically evolved workflows in high-stakes applications risks propagating errors without adequate human oversight.

\newpage

{\small
\bibliographystyle{plain}
\bibliography{main}
}

\clearpage

\appendix

\section{Count-Frequency Experiments}

\subsection{Count-Frequency Experiments}

\subsubsection{Experimental Setup}

\qblead{Task Definition.} Count-Frequency (CF) is a controlled distributed information aggregation task assessing information coverage and global integration in multi-agent workflows. Given an integer array $A$ partitioned into $N$ disjoint private shards $A_1,\ldots,A_N$, each Worker accesses only its own shard. Through inter-agent communication and collaboration, the system must recover the global frequency of each integer:
\qbdisplay{%
c(v)=\sum_{i=1}^{N}\sum_{a\in A_i}\mathbb{I}[a=v],
}
where $v$ belongs to a finite value set $\mathcal{U}$. A prediction is fully correct only if every count in the predicted frequency table is correct.

\qblead{Experimental Configuration.} The original CF experiments use 8 Workers, primarily with array length $|A|=1024$ and integer values 0–9. Core experiments use DeepSeek-V4-Flash for execution, planning, and analysis, with additional exploratory runs using DeepSeek-V3, Qwen3.5, GPT-4o, and MiMo-V2-Flash. The core evolution consists of 5 rounds, each generating 4 candidate structures and retaining 2 for further optimization.

\qblead{Baselines.} We consider three fixed communication topologies: Tree progressively integrates local counts through hierarchical reduction; Mesh–Star exchanges information through dense communication before central aggregation; and One-peer Exponential DAG–Star combines exponentially spaced single-neighbor communication with final star aggregation. Cold-start Free DAG denotes a freely generated communication structure without historical experience, while Self-evolved Free DAG denotes a structure obtained through repeated generation, execution evaluation, and historical candidate retention.

\qblead{Evaluation Metrics.} CF measures the difference between predicted and true frequency tables using:
\qbdisplay{%
\mathrm{RMSE}=\sqrt{\sum_{v\in\mathcal{U}}(\hat{c}(v)-c(v))^2},
}
where $\hat{c}(v)$ and $c(v)$ are the predicted and true counts. Following the original CF implementation, this metric takes the square root of summed squared errors without dividing by the number of distinct values. We also report Messages (inter-agent messages), Model Calls (execution-model calls), and Token Cost (total input and output tokens during execution). Lower values are better for all metrics.

\subsubsection{Main Results}

We compile the original CF results across model configurations to examine free-structure generation under different Worker and Planner settings. Table A2 preserves the originally reported fixed-topology results, selected generated structures, and relative error differences, while adding the actual sample counts for generated results.

\begin{table*}[h]
\centering
\begingroup\small
\begin{adjustbox}{max width=\textwidth}
\begin{tabular}{lrrrr}
\toprule
Model / Configuration & |A| & Best Fixed RMSE ↓ & Selected Graph RMSE ↓ & Reported Reduction \\
\midrule
DeepSeek-V4-Flash & 1024 & 12.53 & 7.87 & 37.2\% \\
DeepSeek-V3 & 1024 & 18.71 & 15.97 & 14.7\% \\
Qwen3.5-Flash + DeepSeek-V4-Pro Planner & 2048 & 21.29 & 14.46 & 32.1\% \\
Qwen3.5-Flash & 1024 & 16.44 & 11.53 & 29.9\% \\
Qwen3.5-Plus & 1024 & 31.16 & 25.94 & 16.7\% \\
GPT-4o (Run 1) & 512 & 14.76 & 6.48 & 56.1\% \\
GPT-4o (Run 2) & 512 & 13.83 & 7.87 & 43.0\% \\
GPT-4o (Run 3) & 512 & 16.05 & 11.10 & 30.8\% \\
MiMo-V2-Flash & 1024 & 43.39 & 37.26 & 14.1\% \\
\bottomrule
\end{tabular}
\end{adjustbox}
\endgroup
\renewcommand{\thetable}{A1}
\caption{Original cross-model CF results.}
\end{table*}

\qblead{Results Analysis.} Across multiple model families, the original experiments generated candidate structures with lower errors than their corresponding fixed-topology references, with reported relative reductions of 14.1\%–56.1\%. These results show that free-structure search can explore diverse information aggregation strategies across model configurations. Overall, the CF experiments demonstrate the exploratory capability of automated communication structure generation and the potential of some candidates for efficient information aggregation.

\subsection{Case Study}

\qblead{Experimental Setup.} We use 8 Subagents to process an integer array of length 1024 with values 0–255, each accessing only 128 integers. The Planner is DeepSeek-V4.1-Flash-thinking-max, and the Worker is gpt-oss-120b. Training, validation, and testing use disjoint random seeds, with final workflows evaluated on 10 independent test instances.

We compare against three fixed topologies (Tree, One-peer Exponential DAG–Star, and Mesh–Star) and a Cold-start Graph without inherited experience. Performance is measured by RMSE between predicted and true frequency tables, with lower values indicating more accurate aggregation. We compare errors on paired test instances and report 95\% bootstrap confidence intervals and two-sided sign-flip test results.

\begin{table*}[t]
\centering
\begingroup\small
\begin{adjustbox}{max width=\textwidth}
\begin{tabular}{lrrrr}
\toprule
Method & Mean RMSE ↓ & Paired ΔRMSE [95\% CI] ↓ & Wins / Losses & p-value \\
\midrule
\textbf{QueenBee (Champion)} & \textbf{3.28 ± 0.84} & — & — & — \\
Tree (Best Fixed) & 5.06 ± 2.45 & −1.78 [−3.01, −0.58] (−35\%) & 8 / 2 & 0.020 \\
One-peer Exponential DAG–Star & 5.42 ± 1.59 & −2.14 [−3.09, −1.12] (−39\%) & 8 / 2 & 0.008 \\
Mesh–Star & 15.55 ± 17.24 & −12.27 [−23.72, −5.37] (−79\%) & 10 / 0 & 0.002 \\
Cold-start Graph & 14.40 ± 16.53 & −11.12 [−21.90, −3.23] (−77\%) & 9 / 1 & 0.006 \\
\bottomrule
\end{tabular}
\end{adjustbox}
\endgroup
\renewcommand{\thetable}{A3}
\caption{Workflow optimization results on Count-Frequency.}
\end{table*}

\qblead{Results Analysis.} As Table A3 shows, QueenBee's evolved champion achieves the lowest mean RMSE (3.28 ± 0.84) across 10 test instances and the smallest error variation, compared with standard deviations of 1.59–17.24 for other methods. Relative to the best fixed topology, Tree, mean RMSE decreases by 35\%, with 8 wins and 2 losses across 10 instances (p = 0.020). Relative to other fixed topologies and Cold-start Graph, errors decrease by 39\%–79\%, with all corresponding p-values below 0.01.

The champion's 95\% bootstrap confidence interval for mean RMSE is [2.78, 3.77], below the mean errors of all baselines. Cold-start Graph has a mean RMSE of 14.40, substantially higher than QueenBee, indicating better aggregation by the experience-guided structure in these tests.

These results demonstrate QueenBee's potential to discover high-quality communication structures for precise distributed information aggregation, providing supplementary evidence of its applicability beyond Silo-Bench.

\section{Qualitative Case Study of Ablation}

\subsection{Comparison of Evolved Workflows}

To further examine how SFT-Bank components affect workflow design, we compare the final executable programs selected by full QueenBee and four ablated variants. All five runs start from the same initial relay workflow but yield distinct communication structures and information-processing instructions.

\begin{table*}[t]
\centering
\begingroup\small
\begin{tabularx}{\textwidth}{ll>{\raggedright\arraybackslash}X}
\toprule
Method & Communication Structure & Information Processing Strategy \\
\midrule
\textbf{QueenBee (Full)} & Gather → Broadcast & Centralized collection and broadcast of complete raw data; independent computation by each Worker \\
w/o Diagnosis & Gather → Broadcast & Centralized computation of the global result and broadcast \\
w/o Ledger & Gather → Broadcast & Centralized computation of the global result; raw data may be omitted \\
w/o Archive & Relay → Broadcast & Retains the relay chain; the final Worker merges information \\
w/o SFT-Bank & Gather → Broadcast & Centralized collection and broadcast without explicit broadcast instructions \\
\bottomrule
\end{tabularx}
\endgroup
\renewcommand{\thetable}{B1}
\caption{Communication structures and information-processing strategies across ablations.}
\end{table*}

Full QueenBee adopts a \qbemph{Complete Snapshot} strategy: Workers first send all private data to a central node, which organizes raw information by source ID and broadcasts it, explicitly forbidding omissions, modifications, reordering, or premature computation of the final answer. In contrast, w/o Diagnosis and w/o Ledger delegate global computation to the central node, w/o Archive retains the initial relay structure, and the memory-free variant lacks explicit constraints on final broadcast content. These differences show that automated workflow search changes not only communication topology but also information preservation and computational responsibilities.

\subsection{Performance on Representative Tasks}

We select three test tasks with substantial performance differences across ablations: I-10 (Standard Deviation), II-14 (1D Life Game), and II-20 (Pipeline Hash). Each final workflow is independently tested three times per task using the original Silo-Bench scoring and unified stream-interruption handling rules.

\begin{table}[t]
\centering
\begingroup\small
\begin{adjustbox}{max width=\columnwidth}
\begin{tabular}{lrrr}
\toprule
Method & I-10 & II-14 & II-20 \\
\midrule
\textbf{QueenBee (Full)} & \textbf{3/3} & 2/3 & \textbf{3/3} \\
w/o Diagnosis & \textbf{3/3} & \textbf{3/3} & 1/3 \\
w/o Ledger & 1/3 & 2/3 & \textbf{3/3} \\
w/o Archive & 1/3 & 0/3 & 2/3 \\
w/o SFT-Bank & 0/3 & \textbf{3/3} & 2/3 \\
\bottomrule
\end{tabular}
\end{adjustbox}
\endgroup
\renewcommand{\thetable}{B2}
\caption{Strict success counts on representative tasks (three runs).}
\end{table}

On I-10, full QueenBee and w/o Diagnosis achieve three fully correct runs, while w/o Ledger and w/o Archive succeed only once each. In one w/o Ledger failure, all Workers submit the same incorrect answer; its centralized computation strategy permits omitting raw data after producing the final result.

On II-20, full QueenBee succeeds in all three runs, while w/o Diagnosis succeeds only once, with dispersed incorrect Worker answers in the other two. II-14 shows a different pattern: w/o Diagnosis and the memory-free variant succeed in all three runs, while full QueenBee has one partially incorrect run. Thus, the benefits of information-processing strategies are task-dependent, and no single task explains overall performance differences across ablations.

\subsection{Evolution of the Complete-Snapshot Strategy}

To trace how full QueenBee's final design emerged during search, we examine its proposal Ledger. On training task II-16 (Trapping Rain), three related structural modifications appeared across successive generations:
\begin{center}
\begingroup\footnotesize
\setlength{\tabcolsep}{2pt}
\begin{tabularx}{\columnwidth}{ll>{\raggedright\arraybackslash}Xlr}
\toprule
Generation & Proposal & Structural Modification & Decision & ΔS \\
\midrule
0 & g000\_s2 & Retain the original structure; add work instructions only & Refuted & −0.2 \\
1 & g001\_s3 & Switch to centralized gathering and broadcasting without explicit instructions & Inconclusive & — \\
2 & g002\_s1 & Centralized gathering and broadcasting with explicit preservation of a complete, ordered data snapshot & Confirmed & +1.0 \\
\bottomrule
\end{tabularx}
\endgroup
\end{center}

The final champion, g002\_s1, targets the historical \textit{scattered-wrong} failure type by proposing a shared information snapshot to prevent Workers from answering based on inconsistent local information. The Planner predicts a 0.75 gain on the target task, while the actual head-to-head evaluation measures a gain of 1.0 (ΔS denotes the change in the fraction of Workers answering correctly).

These records illustrate distinct exploration directions: modifying work instructions alone, changing communication structure, and adding explicit data-integrity constraints. Importantly, the three proposals are not direct parent–child descendants. Thus, the records show that related designs were proposed and evaluated across generations, but do not establish that the final proposal arose through successive modifications of the two earlier failed attempts.

\section{Case Study on Planner-Generated Workflows}

To examine differences in structural design and information-processing rules across Planners, we compare QueenBee champion programs evolved by DeepSeek-V4.1-Flash-thinking-max and gpt-oss-120b-thinking-high. Both runs start from the same relay workflow but use their respective Worker models during evolution. We fix the test Worker to gpt-oss-120b and analyze Silo-Bench task III-29 (Load Balance).

\begin{table*}[t]
\centering
\begingroup\small
\begin{tabularx}{\textwidth}{l>{\raggedright\arraybackslash}X>{\raggedright\arraybackslash}X}
\toprule
Dimension & DeepSeek Planner & gpt-oss Planner \\
\midrule
Champion & g002\_s1 & g005\_s2 \\
Communication & Gather → Broadcast & Relay → Digest → Broadcast \\
Information Preservation & Explicit source labels; complete preservation of raw data and order & General instructions to include received information and private data \\
Computation & Broadcast a complete snapshot; each Worker computes independently & Relay information along the chain, then broadcast \\
III-29 (gpt-oss Worker) & 5/5, 5/5, 4/5 Workers correct & 0/5, 4/5, 5/5 Workers correct \\
\bottomrule
\end{tabularx}
\endgroup
\renewcommand{\thetable}{C1}
\caption{Workflow designs generated by two Planners.}
\end{table*}

\qblead{Workflow Design Differences.} The DeepSeek Planner replaces the initial relay chain with centralized gathering and broadcasting, explicitly instructing Workers to label Agent IDs and preserve all data verbatim without omissions or reordering. The central node then broadcasts a shared, complete information snapshot. In contrast, the gpt-oss Planner retains the relay chain and adds a digest stage. All three communication stages use the same general instruction: "Include all facts you have received so far and your private data in your message." Explicit source labels and data-completeness constraints are absent.

\qblead{Execution Analysis.} III-29 requires five Workers to aggregate 100 distributed tasks and compute load variance, maximum, and minimum values using a prescribed greedy algorithm. In one failed gpt-oss execution, Agent 1 omits all 10 tasks from Agent 0 when forwarding information to Agent 2. Agent 4 then omits 66 accumulated upstream tasks during the final broadcast, sending only its own 24 tasks. Although the recorded message transmission loss is zero, none of the five Workers receives complete data. They access only 34, 56, 68, 90, and 90 tasks, respectively, and submit five distinct incorrect answers.

By contrast, the DeepSeek workflow broadcasts an identical snapshot containing all 100 tasks to every Worker in all three runs. All Workers answer correctly in two runs; in the third, only one Worker fails due to its own computation error. This contrast suggests that explicit information-completeness constraints reduce the risk of content omission in multi-hop communication and help distinguish information propagation errors from final computation errors.

\section{Cost Analysis}

We analyze QueenBee's offline search costs, deployment inference costs, and long-term amortization, and discuss suitable applications. QueenBee is a continuously evolving harness that maintains a deployable incumbent workflow and persistent SFT-Bank, enabling further search as new execution evidence arrives. Challengers must pass a promotion gate before replacing the incumbent. Freezing workflows in experiments ensures independent evaluation on unseen tasks without precluding subsequent evolution.

\subsection{Search and Inference Costs}

Table D1 compares token consumption under the same search-budget setting. Search cost measures tokens consumed by one optimization run, while inference cost measures average tokens per test task after deployment. All values are based on API-reported usage.

\begin{table*}[t]
\centering
\begingroup\small
\begin{adjustbox}{max width=\textwidth}
\begin{tabular}{lrrr}
\toprule
Method & Search Tokens (M) ↓ & Inference Tokens (k/task) ↓ & PCS (\%) ↑ \\
\midrule
\textbf{QueenBee} & 5.477 & 62.6 & \textbf{83.1} \\
GPTSwarm & 10.478 & 113.2 & 75.6 \\
G-Designer & 13.513 & 100.5 & 76.7 \\
MaAS & 14.284 & 60.2 & 61.1 \\
EvoMAS & 5.030 & 34.8 & 68.3 \\
Fixed Protocols (P2P, SFS, Broadcast) & — & 73.5–99.2 & — \\
\bottomrule
\end{tabular}
\end{adjustbox}
\endgroup
\renewcommand{\thetable}{D1}
\caption{Search costs, inference costs, and test performance.}
\end{table*}

QueenBee's search cost is approximately 52\%, 41\%, and 38\% of GPTSwarm, G-Designer, and MaAS, respectively. Its inference cost of 62.6k tokens per task is approximately 45\% and 38\% lower than GPTSwarm and G-Designer, respectively, and below all three fixed protocols. MaAS has comparable inference costs, while EvoMAS has lower search and inference costs; however, their PCS is lower by 22.0 and 14.8 percentage points, respectively. QueenBee achieves the highest PCS with relatively low search and inference costs.

\subsection{Cost Breakdown}

A single QueenBee evolution run consumes 5.477M tokens: approximately 4.863M for task execution and 0.613M for 16 successful Planner calls. Task execution comprises approximately 3.817M tokens for evolutionary evaluation and 1.048M for final validation and selection. Thus, approximately 89\% of search costs arise from task execution, primarily paired challenger–parent comparisons and final validation, while Planner-generated structural modifications account for approximately 11\%.

SFT-Bank management is not a major token cost. For comparison, the memory-free variant consumes 5.550M tokens per search, while removing Diagnosis Cards or the Program Archive yields 6.507M and 7.005M, respectively. Thus, structured experience does not increase total search token consumption in these experiments.

\subsection{Cost Amortization}

QueenBee's main additional costs occur during offline evolution and do not increase per-task search overhead after deployment. Moreover, evaluations of different candidates and training tasks within an iteration are independent and can run in parallel to reduce wall-clock search time.

For systems repeatedly processing similar tasks, offline search costs can be amortized over deployment. QueenBee has lower cumulative token consumption than GPTSwarm and G-Designer from the first deployed task, and amortizes its search cost relative to search-free fixed protocols after approximately 150–505 tasks.

One evolution run costs approximately as many tokens as 88 tasks executed by the final QueenBee workflow. Over 10,000 tasks, its amortized evolution cost is approximately 0.548k tokens per task, or 0.87\% of per-task inference cost.

Thus, QueenBee's cost advantage depends on optimization overhead, deployment efficiency, and task volume, rather than search or inference cost alone.

\subsection{Applicable Scenarios}

QueenBee is particularly suited to multi-agent systems that repeatedly process similar tasks. For example, analyses requiring periodic joint statistics, cross-checking, or auditing across departments or institutions with distributed data can reuse evolved workflows, reducing amortized search costs.

Structural optimization is also valuable when coordination becomes a bottleneck as teams grow. In the 10-Worker Silo-Bench experiment, QueenBee improves PCS by 49.9 percentage points over the initial workflow, indicating its ability to mitigate information integration difficulties in larger teams.

For applications with high error costs or stringent quality requirements, QueenBee can trade additional offline validation for more reliable collaboration structures. Persistent SFT-Bank further enables deployed workflows to improve as new execution evidence accumulates.

Overall, QueenBee trades reusable offline search costs for higher-quality collaborative workflows. Its evolution costs can be amortized through sustained deployment, particularly in applications with recurring tasks, complex coordination, or strict quality requirements.

\section{Additional Method Details}

\subsection{Formal Design Objective}

Rather than directly searching for high-performing workflows, QueenBee accumulates reusable design experience through repeated candidate generation, execution, and evaluation. The design system and its update at iteration $t$ are defined as:
\qbdisplay{%
\begin{gathered}
\mathcal{M}_t=(\mathrm{Planner}_{\theta},\mathcal{B}_t), \qbsepn
\mathcal{P}_t=\operatorname{Generate}(\mathcal{M}_t;b_t), \qbbreak
\mathcal{B}_{t+1}=U\left(\mathcal{B}_t,\mathcal{P}_t,
\operatorname{Eval}(\mathcal{P}_t)\right).
\end{gathered}
}

Here, $\mathrm{Planner}_{\theta}$ is a fixed-parameter design model, $\mathcal{B}_t$ is experience accumulated over $t$ iterations, $b_t$ is the current design budget, $\mathcal{P}_t$ is the candidate set, $\operatorname{Eval}$ denotes execution evaluation, and $U$ updates experience. Updating external experience rather than Planner parameters changes the information and search guidance available for subsequent design.

We measure design capability by the quality of workflows generated under a fixed budget $b$:
\qbdisplay{%
J(\mathcal{M}_t;b)
=
\mathbb{E}_{x\sim\mathcal{D}_{\mathrm{eval}}}
\left[Q(P_t^{*},x)\right],
}
where $P_t^{*}$ is generated within budget $b$ and selected using development-stage evidence, while $Q$ measures its execution quality on an independent evaluation task $x$. Thus, $J$ measures workflow design capability under equal resource constraints, not the performance of an existing workflow. Our objective is:
\qbdisplay{%
\max_{\mathcal{H}_T:\,C_{\mathrm{explore}}(\mathcal{H}_T)\leq B}
J(\mathcal{M}_T;b)
}
where $\mathcal{H}_T$ is a feasible exploration history comprising candidate generation, execution, and feedback; $B$ is the total experience accumulation budget; and $\mathcal{M}_T$ is the design system after $T$ updates.

Thus, we aim to improve workflow search and optimization efficiency under limited budgets through a fixed Planner and evolving external experience, ultimately yielding better workflows.

\subsection{Exploration Scheduling Details}

At iteration $t$, the harness schedules goals and parent workflows for $K$ proposal slots:
\qbdisplay{%
(g_{t,k},P_{t,k}^{\mathrm{par}})
=
\operatorname{Schedule}(\mathcal{B}_t,k).
}

Here, $k$ denotes the proposal slot. Most proposals target quality improvements, while the remainder rotate among structural merging, cost optimization, and open exploration.

For targeted improvements, the harness groups unresolved training tasks by output format and dominant failure type. An Upper Confidence Bound (UCB) strategy prioritizes groups based on remaining improvement potential, historical quality changes, and exploration counts. Directions repeatedly refuted in the Ledger $\mathcal{L}_t$ receive lower priority, while each iteration favors coverage of distinct groups.

For each selected goal, the harness chooses a strong task-specific program from the Program Archive $\mathcal{A}_t$ as the parent, allowing candidates to inherit different local strengths and explore multiple directions.

The harness also periodically schedules three exploration modes: Merge combines workflows with complementary strengths across training tasks; Economize reduces model calls and token consumption while preserving quality; and Explore targets unresolved, underexplored tasks to discover alternatives to existing successful designs.

\subsection{Experience Retrieval Details}

The harness constructs each candidate's design context using the three SFT-Bank components.

From Diagnosis Cards $\mathcal{D}_t$, it categorizes problematic training tasks by output format and observed failure cause, then retrieves the parent's task-specific diagnoses.

From the Ledger $\mathcal{L}_t$, it aggregates validation outcomes and observed performance changes by failure type and modification mechanism, extracting recently supported hypotheses and repeatedly ineffective modifications.

From the Program Archive $\mathcal{A}_t$, it prioritizes successful programs that addressed similar failures and corresponding failure cases, providing complementary programs when combining workflows. Retrieved examples are presented as editable code snippets, allowing the Planner to reference concrete communication structures and execution strategies.

The Planner also receives the parent's editable Python code (Genome), its task-level quality and cost records, the current exploration goal, and fixed execution-interface constraints. The resulting design hypothesis specifies the target problem, modification mechanism, and predicted improvement, while the Genome provides its Python implementation.

\subsection{Validation and Experience Consolidation Details}

Candidates first undergo code and execution-structure validity checks. Duplicate programs, execution-constraint violations, and historically ineffective modifications are rejected before task testing.

In Stage 1, the harness executes candidates on designated target training tasks and reruns their parent workflows under identical conditions. Only candidates satisfying predefined performance requirements without output-format errors advance.

In Stage 2, promising candidates are evaluated on additional training tasks covering diverse templates and output formats. Regression tests check whether previously strong tasks deteriorate. Newly discovered task-specific specialists may undergo independent confirmation tests to reduce the influence of incidental execution outcomes.

After evaluation, the harness derives experience from actual execution results and code changes rather than Planner self-assessment.

Diagnosis Cards retain structured execution statistics by predefined failure type but exclude specific answers and ground-truth labels for safe reuse.

The Ledger identifies added or removed communication stages, round-scheduling adjustments, and changes to work instructions through parent–candidate source comparisons. Based on target-task performance and regression tests, hypotheses are classified as Confirmed, Refuted, or Inconclusive; unexecuted proposals are marked Screened. Confirmation requires a predefined improvement margin without exceeding the allowed regression threshold, while substantial target-task degradation leads to refutation.

The harness links modification mechanisms, validation outcomes, and observed performance changes, tracks historical performance across failure-type–modification combinations, and flags modifications repeatedly refuted in independent tests.

The Program Archive retains new programs, parent-version relationships, training evaluations, and metadata, maintaining historical candidates according to task performance. Screened or insufficiently supported candidates may also leave records; experience updates do not require candidates to outperform existing workflows.

\section{Focused Validation and Final Selection}

When exploration ends and a final workflow is required, the harness stops generating candidates, selects historically strong programs from the Program Archive, and determines the final workflow through independent validation.

To prevent programs with chance high scores on a few tasks from dominating, candidate screening considers both task coverage and overall quality improvement. Specifically, the harness evaluates historical candidates using a shrunk training score relative to the initial workflow $P_0$:
\qbdisplay{%
\widetilde{S}_{T}(P)
=
\frac{1}{|\mathcal{U}_P|+2}
\sum_{u\in\mathcal{U}_P}
\left(
\overline{S}_{P}(u)
-
\overline{S}_{P_0}(u)
\right),
}
where $\mathcal{U}_P$ is the set of tasks with training evaluations for both the candidate and initial workflows, and $\overline{S}_{P}(u)$ is the mean execution quality of program $P$ on task $u$. The denominator's smoothing term penalizes candidates evaluated on too few tasks, preventing locally strong programs from being prematurely ranked as overall best. The harness further requires sufficient training-task coverage and excludes programs showing substantial regression on previously stable tasks, yielding a representative set of high-performing candidates.

The harness then repeatedly evaluates selected programs on identical independent validation tasks, ranks them by mean quality across tasks, and favors lower execution cost when quality is comparable. Top-ranked candidates undergo further regression validation to detect substantial degradation on previously stable tasks. Independent validation results are used solely for final selection and are not fed back to the Planner for further structural modifications, preventing validation information from influencing earlier experience accumulation.

Finally, the harness selects an independently validated executable workflow $P^{*}$. Thus, broad exploration discovers and accumulates diverse communication and collaboration strategies, while focused validation selects a strong overall workflow from historical candidates, completing the process from continual experience accumulation to final workflow selection.

\par\ifdim\dimexpr\pagegoal-\pagetotal\relax<250pt\newpage\fi

\section{QueenBee Results by Difficulty Level}

Table G1 breaks down QueenBee's results under the GPT configuration (Tables H1 and 2) by Silo-Bench difficulty level.

\begin{center}
\begingroup\small
\begin{adjustbox}{max width=\columnwidth}
\begin{tabular}{llrrr}
\toprule
Team size & Difficulty & SR↑ & PCS↑ & Tok.↓ \\
\midrule
N=2 & Level I & 100.0 & 100.0 & 17.3 \\
 & Level II & 25.0 & 37.5 & 16.8 \\
 & Level III & 50.0 & 75.0 & 20.8 \\
N=5 & Level I & 75.0 & 80.0 & 97.9 \\
 & Level II & 25.0 & 45.5 & 99.5 \\
 & Level III & 50.0 & 50.0 & 140.4 \\
N=10 & Level I & 75.0 & 87.5 & 74.9 \\
 & Level II & 0.0 & 38.0 & 61.8 \\
 & Level III & 50.0 & 50.0 & 179.2 \\
\bottomrule
\end{tabular}
\end{adjustbox}
\endgroup
\renewcommand{\thetable}{G1}
\makeatletter\def\@captype{table}\makeatother
\caption{QueenBee by difficulty level across team sizes.}
\end{center}

\par\ifdim\dimexpr\pagegoal-\pagetotal\relax<250pt\newpage\fi

\section{Results with GPT-5.1 and GPT-5 mini}

Table H1 reports the comparison of Table 1 under the GPT configuration (GPT-5.1 Planner, GPT-5 mini Workers). Under GPT, QueenBee achieves 70.8\%, 58.5\%, and 58.5\% PCS with 2, 5, and 10 Workers, exceeding the best fixed protocol, Broadcast, by 1.2, 3.8, and 10.6 percentage points, respectively.

\begin{center}
\begingroup\small
\begin{adjustbox}{max width=\columnwidth}
\begin{tabular}{lccc}
\toprule
Method & $n=2$ & $n=5$ & $n=10$ \\
\midrule
\multicolumn{4}{l}{\textit{Fixed Protocols}} \\
\quad P2P                &  37.5 &  37.4 &  22.3 \\
\quad Broadcast          &  69.6 &  54.7 &  47.9 \\
\quad SFS                &  35.0 &  15.4 &  13.9 \\
\quad Initial Workflow   &  58.3 &  51.4 &   8.6 \\
\textbf{QueenBee (Ours)} & \textbf{70.8} & \textbf{58.5} & \textbf{58.5} \\
\bottomrule
\end{tabular}
\end{adjustbox}
\endgroup
\renewcommand{\thetable}{H1}
\makeatletter\def\@captype{table}\makeatother
\caption{QueenBee versus fixed communication protocols with 2, 5 and 10 Workers under the GPT configuration (GPT-5.1 Planner, GPT-5 mini Workers); fixed protocols use the same Workers.}
\end{center}

\end{document}